# Molecular pillar approach to grow vertical covalent organic framework nanosheets on graphene: new hybrid materials for energy storage

Jinhua Sun,[a] Alexey Klechikov,[a] Calin Moise,[b] Mariana Prodana,[b] Marius Enachescu,[b] and Alexandr V. Talyzin*[a]

**Abstract:** Hybrid 2D-2D materials composed by perpendicularly oriented covalent organic framework (COFs) and graphene were prepared and tested for energy storage applications. Diboronic acid molecules covalently attached to graphene oxide (GO) were used as nucleation sites for directing vertical growth of COF-1 nanosheets (v-COF-GO). The hybrid material shows forest of COF-1 nanosheets with thickness of ~3 to 15 nm in edge-on orientation relative to GO. The same reaction performed in absence of molecular pillars resulted in uncontrollable growth of thick COF-1 platelets parallel to the surface of GO. The v-COF-GO was converted into conductive carbon material preserving the nanostructure of precursor with ultrathin porous carbon nanosheets grafted to graphene in edge-on orientation. It was demonstrated as high-performance electrode material for supercapacitors. The molecular pillar approach can be used for preparation of many other 2D-2D materials with control of their relative orientation.

Hybrid graphene-based composites which combine different types of two dimensional (2D) sheets into one material are some of the hottest research subjects at the moment. Here we explore new hybrid materials of graphene with Covalent Organic Frameworks (COFs). COFs are crystalline polymers obtained using large variety of organic building blocks thus providing hundreds of porous high surface area structures.[1] Combining precise pore size of COFs with high conductivity of graphene in 2D-2D materials could be an advantage for applications in supercapacitors. Furthermore, COFs can be converted into boron doped carbon materials using combination of annealing in molten salts and chemical removal boron oxides.[2]

So far COFs have been almost exclusively studied as microcrystalline powders with only few attempts to prepare single layered nanosheets or very thin nanoplatelets.[3] Moreover, the only available literature reports of graphene/COF composites demonstrated deposition of COF layers or thin COF film *parallel* to the substrate.[3a, 3b, 4] However, the regular packing of COF and graphene layers into parallel stacks will not provide interconnected pore network suitable e.g. for diffusion of ions or sorption of gases. The graphene in such stacks will close the pores of COF making them inaccessible. Here we suggest that structure with theoretically maximal surface area could be composed by COF sheets covalently attached to graphene in edge-on-plane orientation. However, it seems to be technically rather challenging to prepare single layers of COFs and then to attach them to graphene by covalent functionalization since the COFs are not soluble.[1b] Alternatively, we explore possibility to directly grow 2D COF-1 sheets on Graphene to produce covalently attached 2D COF-1 *perpendicular* to graphene. Covalently functionalized graphene oxide (GO) was previously proposed for growth of 3D Metal Organic Framework (MOF) nanocrystals but to our knowledge not for preparation of 2D-2D materials with perpendicular relative orientation.[5]

Here we report a controllable two-step synthesis method which enable growth of ultrathin COF-1 nanosheets oriented vertically on the surface of GO (v-COF-GO). The edge-on anchoring of benzene-1,4-diboronic acid (DBA) molecules to GO determines the growth direction of COF-1 nanosheets approximately perpendicular to the GO surface. Carbonization of v-COF-GO nanocomposites under the protection of molten $ZnCl_2$ allows to produce porous carbon material with 2D nanosheets structure vertically linked on the surface of reduced GO (v-CNS-RGO). The carbon nanocomposites were tested as electrode materials for supercapacitors.

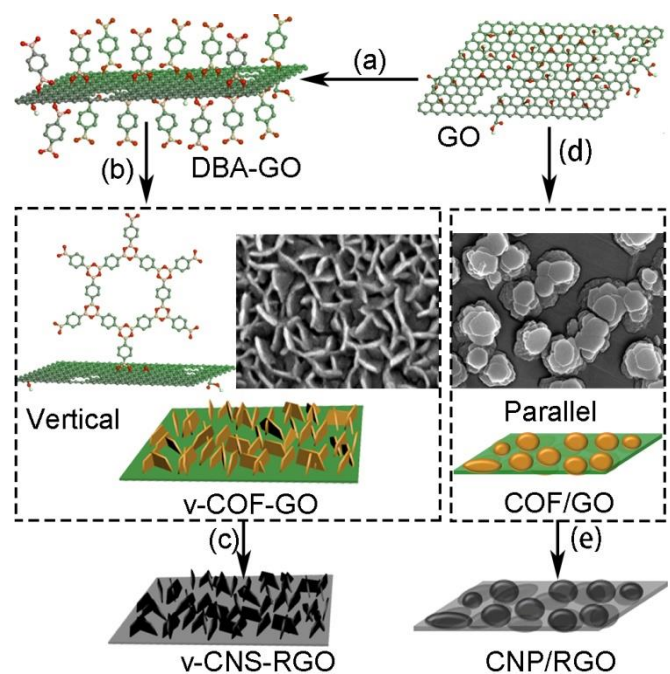

***Scheme 1.*** (a) Covalent functionalization of GO with DBA in methanol solution. (b) Growth of vertical COF-1 nanosheets using DBA as molecular nucleation sites grafted on GO in mesitylene/dioxane solution (v-COF-GO). (d) COF-1 platelets parallel to GO are formed at the same reaction conditions in absence of DBA functionalization (COF/GO). Carbonization of (c) v-COF-GO and (e) COF/GO to produce carbon nanosheets oriented vertical or parallel to RGO surface.

[a] Dr. J.h. Sun, Mr. A. Klechikov, Prof. A. V. Talyzin
Department of Physics
Umeå University
S-90187, Umeå (Sweden)
E-mail: alexandr.talyzin@umu.se
[b] Mr. C. Moise, Dr. M. Prodana, Prof. M. Enachescu
Center for Surface Science and NanoTechnology
University Politechnica of Buchartest
060042 Bucharest (Romania)

Supporting information for this article is given via a link at the end of the document.

The solvothermal reaction of GO in *methanol* solution of DBA[6] results in covalent anchoring of molecular "pillars" to the surface of GO while formation of COF-1 is avoided (Figure S1).[6-7] Since the DBA molecules are attached to GO only by one side,[6] two unreacted -OH groups on the other side can be used as nucleation sites for polycondensation reaction at standard for COF-1 conditions (in mesytilene/dioxane solvent) (Scheme 1b). As a result single layers of COF-1 covalently attached and oriented perpendicular to the graphene surface should be formed. A series of v-COF-GO samples were synthesized using variation of loading ratio of DBA to DBA-GO (1:2, 1:1, 2:1, 3:1, 4:1, and 5:1), named in further discussions as v-COF-GO-x, with x=0.5, 1, 2, 3, 4, or 5.

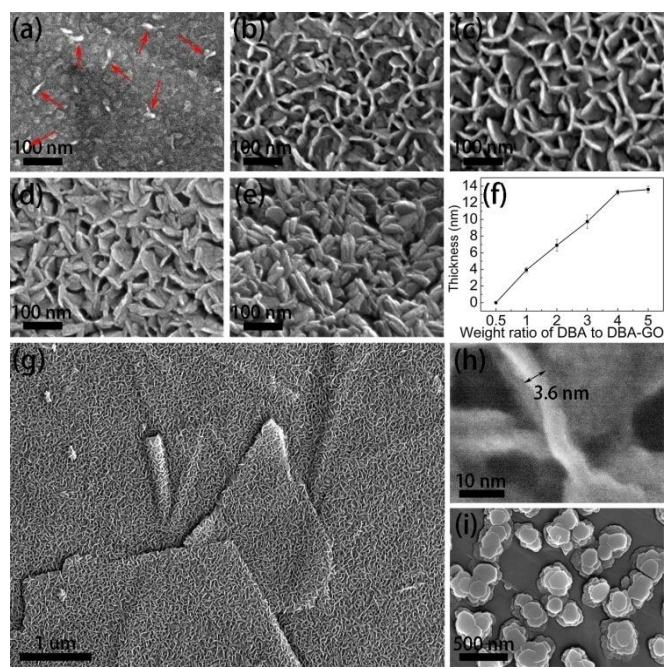

*Figure 1.* SEM images of v-COF-GO nanocomposites as a function of loading ratio of DBA to DBA-GO. Few nanosheets observed for smallest DBA loading of (a) 1:2; densely standing vertical COF-1 nanosheets are found at optimal loading ratio with (b) 1:1 , (c) 2:1 and (d) 3:1  larger loading results in massive formation of COF-1 nanosheets completely covering GO surface (e) 4:1. (f) Correlation of the thickness of vertical COF-1 nanosheets on loading ratio, x. (g) The uniform distribution of vertical COF-1 nanosheets on the surface of GO. (h) The thickness of oriented COF-1 nanosheets measured directly thanks to vertical orientation using high resolution SEM and scanning TEM imaging, (i) Control experiments performed with non-functionalized GO precursor yielded micrometer-sized platelets of COF-1 oriented parallel to GO.

As shown by scanning electron microscopy (SEM) and transmission electron microscopy (TEM) imaging, the COF-1 nanosheets were found to be vertically standing on the surface of GO  (Figure S2 and S3). The weight ratios of DBA to DBA-GO of 1:1, 2:1 and 3:1 were found to be best for formation of nanostructures with vertically standing COF-1 nanosheets covalently linked to GO (Figure 1b-d). The thickness of COF-1 nanosheets correlated with DBA loading (~3 nm to 15 nm, Figure 1h,f). The thickness of  nanosheets was estimated using SEM and STEM images and averaged over several points/nanosheets measurements for each sample in Figure 1 f.

The evidence of COF-1 formation on the surface of GO was also provided using XRD (Figure 2a), Raman spectroscopy and TGA (Figure S4 and S5). The sample with highest initial DBA load and massive crystallization of COF1 crystallites in random orientations showed XRD pattern (additional to GO (001) peak) which is very similar to the pattern of reference COF1 powder sample, both in good agreement with literature data.[1a] However, v-COF-GO obtained with loading ratios of x=1-3 exhibited increased intensity of (100) reflection and decreased intensity of other reflections, most notably (110) reflection compared with as-synthesized COF-1 powder and COF/GO (Figure 2b). The absence of reflections related to out of plane lattice is an evidence of preferential orientation, while similarity of patterns to those typical for AA stacking could be considered as an indication of ultra-small thickness of COF-1 sheets which enables partial evaporation of guest solvent molecules. Specific surface area (SSA) and pore size distribution of the COF-GO nanocomposites were characterized using analysis of $N_2$ adsorption/desorption isotherms (Figure S6 and S7). As expected, SSA of v-COF-GO increases with increasing the content of COF-1 up to ~700 $m^2\,g^{-1}$ (Figure 2c). Interestingly, a large fraction of pore volume was found to originate from mesopores (up to~17%) whereas pure COF-1 is nanoporous material with negligible mesopore volume (Figure 2d and S8).

Control experiments were also performed at the same solvothermal reaction conditions with precursor GO not subjected to DBA functionalization. Absence of covalently attached DBA molecules on the surface of GO resulted in uncontrolled precipitation of flat COF-1 nanoplatelets oriented mostly parallel to GO surface (Figure 1i and Figure S9-S12) and typically of much larger in size (~ 500 nm). These composites will be named as COF/GO to emphasize absence of covalent bonding between COF-1 and GO.

It can be concluded that high density DBA pillars anchored to GO surface provide a key role to oriented growth of ultrathin COF-1 nanosheets.

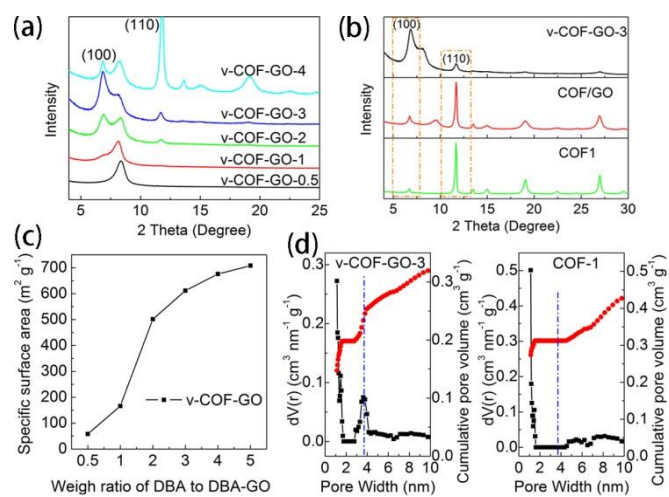

*Figure 2.* (a) XRD patterns of v-COF-GO nanocomposites with different loading ratio of DBA to DBA-GO. (b) XRD patterns of v-COF-GO-3, control sample COF/GO-3 prepared using non-functionalized GO and reference pattern of COF-1 microcrystalline powders. (c) SSA of v-COF-GO as a function of different loading ratio of DBA to DBA-GO. (d) Pore size distribution and cumulative pore volume of v-COF-GO-3 and COF-1 nanocomposites.

Remarkably, the orientation of COF-1 relative to graphene can be preserved even after carbonization in molten salts (see details of procedure in SI). The resulting carbon nanostructures inherited the geometry of precursor COF-1 on GO, forming oriented porous carbon nanosheets either perpendicular or parallel to RGO surfaces (Figure 3 and S13). In contrast, heat treatment of v-COF-GO without molten salt resulted in collapse of vertically attached COF-1 nanosheets (Figure S14).

XRD and Raman spectroscopy data confirmed successful transformation of v-COF-GO structures into amorphous carbon (Figure S15 and S16). The thickness of carbon nanosheets decreased by ~1.5 nm compared to the thickness of precursor COF-1 nanosheets (Figure S17). Trace amounts of boron revealed in the carbonized samples by X-ray photoelectron spectroscopy (XPS) (e.g. 0.51% for v-CNS-RGO-3) (Figure S18) could be considered as an advantage for applications in supercapacitors. It could contribute to pseudocapacitance, and as a result, to improve overall capacitance of supercapacitor electrodes.[8]

The surface area of carbonized v-COF-GO (~500 m$^2$ g$^{-1}$) appeared to be almost as high as for the precursor v-COF-GO (Figure S19). Also the combination of micro- and mesoporosity of COF-GO preserved in carbonized materials (Figure S20). In contrast, the control samples COF/GO with COF-1 nanoplatelets parallel to GO surface provided lower SSA after carbonization (CNP/RGO) (up to ~350 m$^2$ g$^{-1}$, Figure S19).

The performance of v-CNS-RGO nanocomposites as electrodes for supercapacitors was investigated using standard two-electrode electrochemical cell in a 6M KOH electrolyte.[9] The cyclic voltammetry (CV) curves of v-CNS-RGO electrodes at various scan rates showed quasi-rectangular shape, indicating typical capacitive behavior (Figure S21). A big hump observed in CV curves of v-CNS-RGO is a sign of pseudocapacitance which is found to correlate with increased amount of B-doped carbon nanosheets (Figure 4a). The v-CNS-RGO-3 electrode shows largest areas of CV curve and highest intensity of hump as compared with other v-CNS-RGO electrodes, implying the best electrochemical performance. Galvanostatic charge/discharge (GCD) curves of v-CNS-RGO electrodes at different current densities show good symmetry and nearly linear slope (Figure S22).

The v-CNS-RGO-3 electrode showed the highest specific capacitance with good capacitance retention (Figure 4b), comparable to that of reported activated carbon based electrodes.[10] The best specific capacitance of v-CNS-RGO-3 electrode is assigned to combination of high surface area, higher amount of B-doping and unique geometry of material with porous carbon nanosheets vertically grafted to RGO. Too high DBA loading (x=4) results in lower capacitance due to massive deposition of chaotically oriented COF-1 (Figure S23) which makes pores of the structure less accessible to electrolyte.

The electrochemical impedance spectroscopy (EIS) analysis provided Nyquist plots of v-CNS-RGO electrodes (Figure S24).[11] The smallest semicircle was recorded for v-CNS-RGO-3 electrode, indicating a lower charge transfer resistance, thus confirming the fast electron transfer from porous carbon nanosheets to high conductive graphene, and faster ion diffusion.[11-12] The v-CNS-RGO-3 electrode exhibits also excellent cycling stability, without capacitance decay after 3000 cycles at a current density of 10 A g$^{-1}$ (Figure 4d).

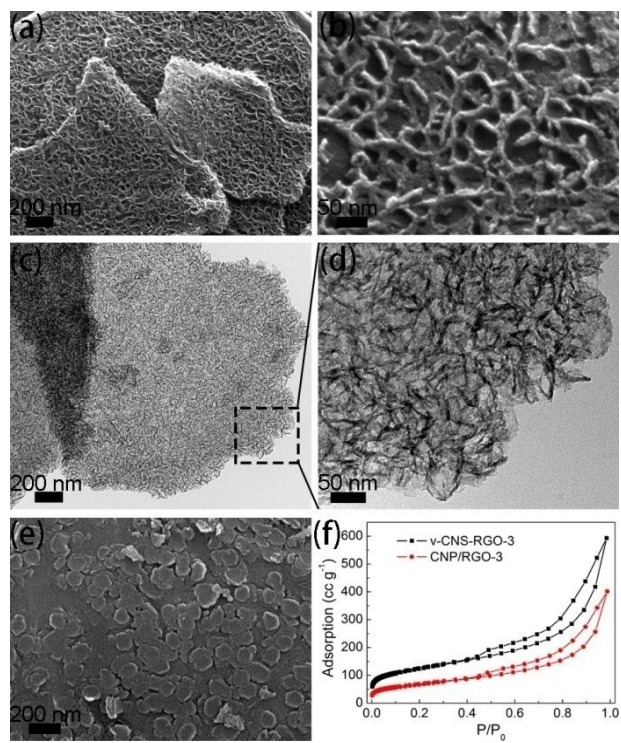

*Figure 3.* (a) SEM and (b) magnified SEM images of v-CNS-RGO-2 nanocomposites. (c) TEM and (d) magnified TEM images of v-CNS-RGO-3 nanocomposites. (e) SEM image of CNP/RGO nanocomposites. (f) Nitrogen sorption isotherms of v-CNS-RGO-3 and CNP/RGO-3 nanocomposites.

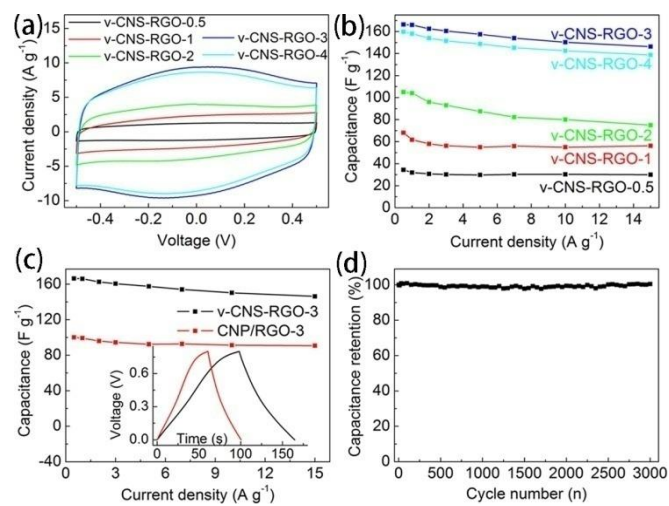

*Figure 4.* (a) CV curves of v-CNS-RGO electrodes prepared with different ratio of DBA to DBA-GO (100 mV s$^{-1}$). (b) Capacitance for v-CNS-RGO electrodes as a function of loading ratio. (c) Capacitance for carbonized samples with vertical (v-CNS-RGO-3) and parallel (CNP/RGO-3) orientation of nanosheets. Inset: GCD curves of v-CNS-RGO-3 and CNP/RGO-3 electrodes at 1 A g$^{-1}$. (d) Cycling performance of the v-CNS-RGO-3 electrode.

The vertical geometry of carbon nanosheets attached to graphene is crucial for achieving best electrochemical performance. The control samples with carbon nanoplatelets parallel to graphene surface obtained using the same loading ratio of DBA to DBA-GO demonstrated significantly lower specific capacitance (Figure 4c), worse electron transfer and ion diffusion (Figure S25-S27).

Materials with high surface area could also be of interest for applications in hydrogen storage. Therefore we evaluated also hydrogen sorption properties of selected v-COF-GO with highest surface area (Figure S28). As expected from our earlier results,[13] precise correlation of hydrogen uptake with SSA was observed both at room temperature and 77K.

In summary, we produced new hybrid nanomaterial with orientation of ultrathin COF-1 nanosheets perpendicular to graphene surface. The DBA molecules covalently attached to the surface of GO in vertical orientation acted as nucleation agent and directed the vertical growth of COF-1 nanosheets. The thickness of COF-1 nanosheets was precisely controlled from ~ 3 to 15 nm (several to tens layers of COF-1) by varying the loading of DBA. Moreover, we demonstrated that GO-2D COF hybrids can be converted into boron doped carbon materials which preserve the unique geometry of their precursor nanostructures. The v-CNS-RGO was demonstrated as high performance electrode material for supercapacitor due to its unique nanostructure which allowed fast electrons transfer from vertical porous carbon nanosheets to highly conductive RGO. Controlling nucleation of 2D sheets and their growth direction by using molecular building blocks covalently attached to graphene as structure-directing agent provides very promising route for preparation of new family hybrid materials.

## Experimental Section

Experimental Details are included in the Supporting Information.

## Acknowledgements

J.S., A.K. and A.T acknowledge funding from the European Union's Horizon 2020 research and innovation program under grant agreement No 696656; support by the facilities and technical assistance of Cheng C. N. at the UCEM at the Chemical Biological Centre (KBC), Vibrational Spectroscopy Platform Umeå University and A.Shchukarev for assistance with XPS. A.T. acknowledges financial support from Carl Tryggers Stiftelse. The authors also acknowledge supported financially by Romanian Ministry of Research and Innovation and by Executive Agency for Higher Education, Research, Development and Innovation, under M-ERA.net Project 37/2016 and ENIAC 04/2014, and also COST Action CA 15107 "Multi-Functional Nano-Carbon Composite Materials Network (MultiComp).